\documentclass[english]{article}
\usepackage[T1]{fontenc}
\usepackage[latin9]{inputenc}
\usepackage[letterpaper]{geometry}
\usepackage{amssymb}
\usepackage{graphicx}
\makeatletter
\providecommand{\tabularnewline}{\\}
\usepackage{setspace}
\usepackage[bf,center]{caption}

\newenvironment{myindentpar}[2]%
{\begin{list}{}%
         {\setlength{\leftmargin}{#1}\setlength{\rightmargin}{#2}}%
         \item[]%
}
{\end{list}}

\makeatletter
\renewcommand\@biblabel[1]{#1.}
\makeatother

\usepackage[numbers]{natbib}
\makeatother

\usepackage{babel}
\begin{document}
\begin{center}
~
\par\end{center}

\begin{center}
\vspace{-0.6cm}

\par\end{center}

\begin{center}
\begin{spacing}{2}\textbf{\LARGE Streaking At High Energies}\\
\textbf{\LARGE{} With Electrons And Positrons}\end{spacing}
\par\end{center}

\begin{center}
\vspace{-1.1cm}

\par\end{center}

\begin{center}
{\large Andreas Ipp$^{\mathrm{a}}$, Jörg Evers$^{\mathrm{b}}$, Christoph
H. Keitel$^{\mathrm{b}}$, Karen Z. Hatsagortsyan$^{\mathrm{b}}$}
\par\end{center}{\large \par}

\vspace{-0.1cm}

\begin{center}
\textit{$^{\mathrm{a}}$Institut für Theoretische Physik, Technische
Universität Wien, 1040 Vienna, Austria. }\\
\textit{$^{b}$Max-Planck-Institut für Kernphysik, Saupfercheckweg
1, D-69117 Heidelberg, Germany}
\par\end{center}

\vspace{-0.1cm}

\begin{myindentpar}{0.51cm}{0.51cm}

\fontsize{10}{11} \selectfont \textbf{\small Abstract.}{\small{}
State-of-the-art attosecond metrology deals with the detection and
characterization of photon pulses with typical energies up to the
hundreds of eV and time resolution of several tens of attoseconds.
Such short pulses are used for example to control the motion of electrons
on the atomic scale or to measure inner-shell atomic dynamics. The
next challenge of time-resolving the inner-nuclear dynamics, transient
meson states and resonances requires photon pulses below attosecond
duration and with energies exceeding the MeV scale. }\\
{\small Here we discuss a detection scheme for time-resolving high-energy
gamma ray pulses down to the zeptosecond timescale. The scheme is
based on the concept of attosecond streak imaging, but instead of
conversion of photons into electrons in a nonlinear medium, the high-energy
process of electron-positron pair creation is utilized. These pairs
are produced in vacuum through the collision of a test pulse to be
characterized with an intense laser pulse, and they acquire additional
energy and momentum depending on their phase in the streaking pulse
at the moment of production. A coincidence measurement of the electron
and positron momenta after the interaction provides information on
the pair production phase within the streaking pulse. We examine the
limitations imposed by quantum radiation reaction in multiphoton Compton
scattering on this detection scheme, and discuss other necessary conditions
to render the scheme feasible in the upcoming Extreme Light Infrastructure
(ELI) laser facility.}{\small \par}

\textbf{\small PACS:}{\small{} 42.65.Re, 07.85.Fv, 41.75.Ht.}{\small \par}

\end{myindentpar}

\vspace{0.1cm}

\begin{center}
\textbf{\large INTRODUCTION}
\par\end{center}{\large \par}

\vspace{0.1cm}

\fontsize{12}{14} \selectfont 

\renewcommand\figurename{FIGURE}

\renewcommand\tablename{TABLE}

\label{intro} The Extreme Light Infrastructure (ELI) wants to break
new ground in many respects. Among other goals, it strives to achieve
the highest intensities and the shortest pulse durations. In fact,
these two aims are related, as has been observed in the so-called
intensity pulse duration conjecture \cite{Mourou}. The shortest possible
pulses are expected to be generated from highest intensity facilities
like ELI \cite{ELIweb} or IZEST \cite{IZEST}. Pulse durations down
to the zeptosecond regime seem reachable. Suggestions to produce zeptosecond
pulses of keV-energy photons include relativistic laser-plasma interactions~\cite{Pukhov,Bulanov,Nomura}.
Short pulses of multi-MeV energy photons can be produced via nonlinear
Thomson/Compton backscattering~\cite{Hartemann,lan:066501,Kim}.
At even shorter timescales, there is a proposal for an imploding ultrarelativistic
flying mirror which can be created by a megajoule energy laser pulse
at the ultrarelativistic intensity of $10^{24}$ W/cm$^{2}$~\cite{Tajima}.
This would be capable of back-scattering a 10-keV coherent x-ray pulse
into a coherent $\gamma$-ray pulse with a duration of 100 ys. Moreover,
double pulses of yoctosecond duration of GeV photon energy could be
created in non-central heavy ion collisions~\cite{Ipp:2009ja,Ipp:2007ng}.

So far, a time-dependent characterization of $\gamma$-ray pulses
in the MeV--GeV energy range is not available yet, even at moderately
short fs-as timescales. An accurate measurement of photon pulses emitted
from extreme laser field driven plasmas, nuclei, or heavy ion collisions
would provide valuable information on the underling physical processes.
At lower energy scales, there exists a variety of methods for attosecond
time resolution. Autocorrelation schemes use the test pulse and its
time-shifted replica (Frequency-Resolved Optical Gating -- FROG~\cite{FROG,Mairesse2005})
or the time- and frequency-shifted replica (Spectral Phase Interferometry
for Direct Electric field Reconstruction -- SPIDER~\cite{Iaconis1998,Quere}),
while cross-correlation schemes are based on the correlation between
the test XUV pulse and a femtosecond infrared laser pulse. The latter
can be weak, inducing few photon effects (Reconstruction of Attosecond
Beating By Interference of Two-photon Transitions -- RABBITT~\cite{Paul2001})
or strong, yielding attosecond streak imaging \cite{Drescher2001,Itatani2002,Kitzler2002}. 

In streak imaging~\cite{Drescher2001} a short test pulse (TP) to
be characterized is co-propagated with an auxiliary streaking pulse
(SP). In the presence of the SP, photons from the TP are converted
to electrons through a nonlinear mechanism. The phase of the SP at
the moment of the electron emission determines its final momentum.
Therefore, the final momentum distribution of the photoelectrons provides
information on the duration and the chirp of the TP. Atomic photoionization
is used for TP energies below 100 eV. In the hard x-ray domain, streak
cameras can be based on Compton ionization~\cite{Yudin}. 
\begin{figure}[b]
\centering \includegraphics[width=0.5\columnwidth]{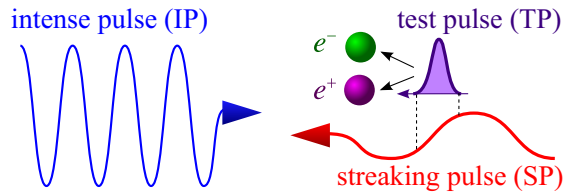}

\caption{\label{fig:1}Concept of SHEEP. Electron-positron pairs are produced
through the interaction of a short test pulse with an intense anti-aligned
laser field within a streaking laser pulse. The leptons acquire additional
energy and momentum depending on their phase in the streaking pulse
at the moment of production. Coincidence measurement of an electron-positron
pair allows the reconstruction of the phase of the streaking pulse
at the time of pair production.}
\end{figure}
 However, for short pulses of $\gamma$-rays exceeding the MeV range~\cite{Landau},
Compton ionization becomes inefficient. New schemes are therefore
required that can characterize the pulses in the sub-attosecond and/or
super-MeV regime that are expected at the projected ELI, HiPER (High
Power laser Energy Research), or IZEST facilities.

In Ref.~\cite{Ipp:2010vk}, we proposed a detection scheme, called
``Streaking at High Energies with Electrons and Positrons'' (SHEEP).
It can be used to characterize short $\gamma$-ray pulses of super-MeV
energy photons down to the zeptosecond scale. Figure~\ref{fig:1}
presents the basic concept of SHEEP. The underlying mechanism of SHEEP
is pair creation from the vacuum: when a $\gamma$-photon collides
with a strong laser beam, an electron and positron pair can be created
due to the absorption of the high-energy $\gamma$-photon and numerous
less energetic photons of the laser beam. The proof-of-principle of
the process at the threshold of the nonperturbative regime has been
provided in the benchmark SLAC experiment E-144 ~\cite{PhysRevLett.79.1626}.

The electron-positron pairs are produced by the interaction of the
TP with a counter-propagating intense laser pulse (IP). Thus, the
IP replaces a photoionization process or Compton ionization from conventional
streak imaging. But different from conventional streak imaging, in
SHEEP two particles with opposite charges, electron and positron,
are created in the same relative phase within a SP that co-propagates
with the TP. By performing a coincidence measurement of the momentum
and energy of electrons and positrons originating from different positions
within the TP, its length and, in principle, its shape can be reconstructed.

\vspace{0.1cm}

\begin{center}
\textbf{\large THE REQUIREMENTS OF SHEEP}
\par\end{center}{\large \par}

\vspace{0.1cm}

\label{concept}

Three photon pulses with specific functions are required for the SHEEP
concept (see Fig.~\ref{fig:1}). The SP that co-propagates with the
TP is linearly polarized, and the polarization of the IP is assumed
to be perpendicular to the one of the SP.

The first requirement for a successful operation of SHEEP is that
a sufficient number of electron-positron pairs is created by the laser
fields. The strong field pair production process is governed by two
relativistic invariant parameters $\xi=e\sqrt{A_{\mu}A^{\mu}}/m$
and $\chi=e\sqrt{(F_{\mu\nu}k_{t}^{\nu})^{2}}/m^{3}$~\cite{Ritus},
with $A_{\mu}$ and $F_{\mu\nu}$ the vector potential and the field
tensor of the laser fields, respectively, $k_{t}$ the TP $4$-momentum,
and $e$ and $m$ the absolute value of the charge and the mass of
the electron. In the chosen geometry $\chi=(k_{i}k_{t})\xi_{i}/m^{2}=2\omega_{i}\omega_{t}\xi_{i}/m^{2}$
and $\xi^{2}=\xi_{i}^{2}+\xi_{s}^{2}$, where $\omega$ denotes the
photon energy, and the indices ``$t$'', ``$s$'', or ``$i$''
refer to TP, SP, or IP, respectively ($\hbar=c=1$). All pairs with
any initial momenta will be analyzed in SHEEP and provide information
on the creation phase in the SP, and thus on the duration of the pulse.
Assuming that all electrons and positrons can be matched correctly,
the initial momenta can in principle be fully reconstructed. 

The second requirement is that the pair production should be initiated
only by $\gamma$-photons of the TP but not by the SP and the IP.
For the latter, first of all, the $\chi_{s}$ parameter associated
with the SP photons should be small $\chi_{s}\equiv2\omega_{i}\omega_{s}\xi_{i}/m^{2}\ll1$.
Moreover, the fields of the SP and IP in the center-of-mass frame
of the electron-positron pairs, hypothetically produced via the SP
and IP, should be negligible with respect to the Schwinger critical
field $E_{cr}=m^{2}/e$ ~\cite{Ritus}. The center-of-mass frame
is determined by the equality of the Doppler-shifted frequencies of
the SP and IP, $2\gamma_{cm}\omega_{i}=\omega_{s}/2\gamma_{cm}$,
with the Lorentz-factor of the center-of-mass frame $\gamma_{cm}$.
The conditions for the suppression of the pair production by the SP
and IP interaction then yield $\sqrt{\omega_{i}\omega_{s}}\xi_{i,s}\ll m\,$.

The electron and positron arise from vacuum in a certain phase of
the SP and move afterwards in the combined field of the IP and SP.
In order to be able to reconstruct the initial phase of the SP in
which the electron-positron pair is created, one needs to demand a
third condition: the electron momentum is far from the resonance condition
corresponding to the stimulated Compton process driven by the SP and
the IP. The off-resonance condition, taking into account the above-threshold
processes, is $\omega_{i}\omega_{t}^{2}\gg2\omega_{s}\xi_{i}^{2}m^{2}$.
This inequality also covers possible sub-threshold processes due to
short pulse effects~\cite{Heinzl}.\pagebreak{}

\vspace{0.1cm}

\begin{center}
\textbf{\large THE RESOLUTION}
\par\end{center}{\large \par}

\vspace{0.1cm}

\label{resolution}

The achievable resolution of SHEEP can be estimated from the energy
and momentum gain of the electron or positron during the motion in
the superposition of the IP and SP. This can be calculated using relativistic
classical equations of motion. The influence of the TP on the electron
motion is negligible during the streaking phase.

Additional information is obtained by measuring the positron energy
in addition to that of the electron energy. The transversal momenta
of electrons and positrons should match, so this can be used to select
corresponding pairs and for a consistency check. The coincidence measurement
of the electron and positron momenta after the interaction provides
information on the pair production phase $\zeta_{0}$ in the SP. Simultaneously,
the measurement determines the emission angles $(\theta,\phi)$ of
the produced electron and the number of photons absorbed during the
process. Therefore the SHEEP measurement determines not only the phase
of the emission $\zeta_{0}$ but also the electron and positron initial
momenta at the creation moment. \begingroup 
\begin{table}
\begin{centering}
\begin{tabular}{clccccc}
\hline 
 &  & \multicolumn{3}{c}{High energy TP} & \multicolumn{2}{c}{Low energy TP}\tabularnewline
 &  & Femto-  & Atto-  & Zeptosecond  & Atto-  & Zeptosecond \tabularnewline
\hline 
IP & $\begin{array}{c}
\\
\omega_{i}\:[\mathrm{eV}]\\
I_{i}\:[\mathrm{W}/\mathrm{cm}^{2}]\\
\xi_{i}\\
{\mathcal{N}}_{i}
\end{array}$  & $\begin{array}{c}
\\
1\,\\
10^{20}\\
10\\
\sim3
\end{array}$  & $\begin{array}{c}
\\
1\,\\
10^{20}\\
10\\
\sim3
\end{array}$  & $\begin{array}{c}
\\
1\,\\
10^{20}\\
10\\
\sim3
\end{array}$  & $\begin{array}{c}
\\
1000\\
10^{24}\\
1\\
\sim30
\end{array}$  & $\begin{array}{c}
\\
1000\\
10^{24}\\
1\\
\sim30
\end{array}$ \tabularnewline
\hline 
SP & $\begin{array}{c}
\\
\omega_{s}\:[\mathrm{eV}]\\
I_{s}\:[\mathrm{W}/\mathrm{cm}^{2}]\\
\xi_{s}
\end{array}$  & $\begin{array}{c}
\\
1\\
10^{18}\\
1
\end{array}$  & $\begin{array}{c}
\\
100\\
10^{22}\\
1
\end{array}$  & $\begin{array}{c}
\\
1000\\
10^{24}\\
1
\end{array}$  & $\begin{array}{c}
\\
100\\
10^{20}\\
0.1
\end{array}$  & $\begin{array}{c}
\\
1000\\
10^{22}\\
0.1
\end{array}$ \tabularnewline
\hline 
TP & $\begin{array}{c}
\\
\omega_{t}\:[\mathrm{GeV}]\\
\tau_{t}\:[as]
\end{array}$  & $\begin{array}{c}
\\
>30\\
10^{2}-10^{3}
\end{array}$  & $\begin{array}{c}
\\
>30\\
1-10
\end{array}$  & $\begin{array}{c}
\\
>30\\
0.1-1
\end{array}$  & $\begin{array}{c}
\\
>0.3\,\\
1-10
\end{array}$  & $\begin{array}{c}
\\
>0.3\,\\
0.1-1
\end{array}$ \tabularnewline
\hline 
\end{tabular}

\par\end{centering}

\caption{SHEEP parameters for different combinations of intense laser sources.
$\Delta\omega_{t}/\omega_{t}\lesssim0.1$, and $N/S=10^{-2}$ are
assumed. $(N_{e+e-}/N_{t})|_{\omega_{t}=\omega_{t\, min}}\sim10^{-2}$
in all cases~\cite{Ipp:2010vk}. \label{tab:comparison}}
\end{table}

\endgroup 

The SHEEP resolution can be estimated from the energy difference $\Delta\mathcal{E}$
of two electrons created at two different pair production phases $\zeta_{1}$
and $\zeta_{2}$, 
\begin{equation}
\Delta\mathcal{E}\sim\omega_{t}\omega_{s}\tau_{t}\max\left\{ \frac{\xi_{s}}{\sqrt{2}\xi_{i}},\frac{\xi_{s}^{2}}{\xi_{i}^{2}}\right\} .\label{deltaE}
\end{equation}
 The energy difference $\Delta\mathcal{E}$ due to streaking should
exceed the energy uncertainty of the TP $\Delta\mathcal{E}\gg1/\tau_{t}$
as well as the bandwidth $\Delta\omega_{t}$ of the $\gamma$-ray
beam $\Delta\mathcal{E}\gg\Delta\omega_{t}$. 

Another effect of potential influence is radiation reaction \cite{DiPiazza:2009zz,DiPiazza:2011tq}.
The electrons and positrons moving in a strong laser field can radiate
via multiphoton Compton scattering. This may modify the electron dynamics
and disturb the SHEEP operation. Radiation reaction will be significant
if the energy loss of the electron due to radiation during the motion
in one laser period is comparable to the initial electron energy.
Inclusion of the radiation reaction in the quantum regime increases
the spectral yield of multiphoton Compton scattering at low energies,
and decreases the spectral yield at high energies \cite{Hatsagortsyan:2011wh}.
However, the probability of a photon emission in the multiphoton Compton
process $W_{C}\sim\alpha\xi_{i}\mathcal{N}_{i}$ will be negligible
when $\alpha\xi_{i}\mathcal{N}_{i}\ll1,$ with the number of cycles
in the IP $\mathcal{N}_{i}$. In the streaking regime we have $\chi\sim1$,
and thus $\alpha\xi_{i}\chi\ll1$. Only in the opposite limit $\alpha\xi_{i}\chi\gtrsim1$,
the radiation dominated regime of multiphoton Compton scattering is
entered~\cite{BulanovRDR,dipRR}. Similarly, a cascade of pair production
\cite{Kirk, Narozhny} can only be initiated for $\chi\gtrsim1$ if
the interaction time $\tau_{i}=2\pi\mathcal{N}_{i}/\omega_{i}$ is
much larger than the pair creation time $\tau_{e^{+}e^{-}}\sim\omega_{t}/\alpha m^{2}\chi^{2/3}$
\cite{Ritus}, which yields $\alpha\xi_{i}\mathcal{N}_{i}/\chi^{1/3}\gg1$.
In the streaking regime, the opposite condition is fulfilled and the
pair production cascade is suppressed. 

Finally, basic preconditions for streak imaging are that the TP length
$\tau_{t}$ is shorter than half of the SP wavelength $\lambda_{s}=2\pi/\omega_{s}$,
and that the streaking signal exceeds the noise level~\cite{Krausz2009},
$\pi N/S\ll\omega_{s}\tau_{t}<\pi,$ where $S/N$ is the signal-to-noise
ratio for the laser fields. The resolution of the TP duration is directly
related to the SP frequency via this condition.

\begin{figure}
\centering \includegraphics[width=0.7\columnwidth]{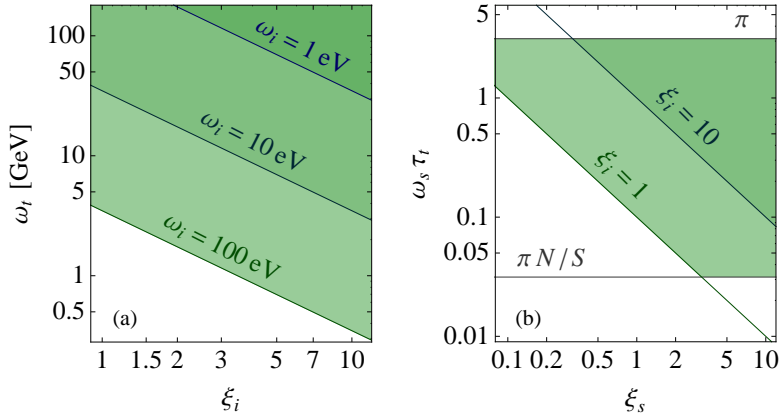} \caption{Possible SHEEP ranges of (a) the TP photon energy and (b) the TP duration
for $\Delta\omega_{t}/\omega_{t}=0.1$ and $N/S=10^{-2}$. The allowed
range of $\omega_{t}$ in (a) is indicated with $\omega_{i}$-dependent
hue, and the range of $\tau_{t}$ in (b) is indicated with $\xi_{i}$-dependent
hue \cite{Ipp:2010vk}. }

\label{fig:2} 
\end{figure}

\vspace{0.1cm}

\begin{center}
\textbf{\large SHEEP PARAMETERS}
\par\end{center}{\large \par}

\vspace{0.1cm}

\label{parameters}

Table \ref{tab:comparison} shows a comparison of different possibilities
to realize SHEEP. The IP is a short and relatively strong laser field
with $\xi_{i}\sim1-10$ and ${\mathcal{N}}_{i}=3-30$. The required
infrared IP with an intensity of $10^{20}$ W/cm$^{2}$ is routinely
available in many labs. The intense high-frequency SP/IP with photon
energies in the $0.1-1$ keV range can be produced in the ELI facility
via high-order harmonic generation at plasma surfaces~\cite{ELIweb}.
Streaking requires detection of at least two electrons emitted from
two different points in time within the TP. As Table \ref{tab:comparison}
shows, this is possible with hundreds of photons per TP.

\vspace{0.1cm}

\begin{center}
\textbf{\large CONCLUSION}
\par\end{center}{\large \par}

\vspace{0.1cm}

\label{conclusion}

SHEEP provides a detection scheme suitable for the characterization
of short $\gamma$-ray pulses in the super-MeV energy range. It is
based on the process of vacuum pair creation in a strong field and
requires a setup of three beams: a strong infrared beam that provides
the necessary intensity for pair creation, an x-ray beam that acts
as a streaking background, and a $\gamma$-ray beam that shall be
characterized. Using high-order harmonic generation in the upcoming
ELI facility, sub-attosecond time resolution could be achieved.

\vspace{0.1cm}

\begin{center}
\textbf{\large ACKNOWLEDGMENTS}
\par\end{center}{\large \par}

\vspace{0.1cm}

\fontsize{12}{14} \selectfont A.~I.~would like to thank the organizers
for an inspiring and exciting conference.

\vspace{0.1cm}

\begin{center}
\textbf{\large REFERENCES}
\par\end{center}{\large \par}

\vspace{-1.4cm}

\fontsize{10}{11} \selectfont \renewcommand\refname{}

\bibliographystyle{unsrt}
\bibliography{ipp_bibliography2,lei2011_bibliography}

\end{document}